\documentclass[a4paper]{jpconf} 

\usepackage{graphicx}
\usepackage{amsmath, amssymb}
\usepackage{hyperref,iopams}
\usepackage{ulem}

\begin{document}

\title{Nuclear physics inputs for dense-matter modelling in neutron stars. The nuclear equation of state}

\author{Anthea Francesca Fantina$^1$ and Francesca Gulminelli$^2$}

\address{$^1$ Grand Acc\'el\'erateur National d'Ions Lourds (GANIL),  CEA/DRF - CNRS/IN2P3,  Boulevard Henri Becquerel,  14076 Caen,  France \\
$^2$ Normandie Univ., ENSICAEN, UNICAEN, CNRS/IN2P3, LPC Caen, 14000 Caen, France}

\ead{anthea.fantina@ganil.fr}

\begin{abstract}
In this contribution, we briefly present the equation-of-state modelling for application to neutron stars and discuss current constraints coming from nuclear physics theory and experiments.
To assess the impact of model uncertainties, we employ a nucleonic meta-modelling approach and perform a Bayesian analysis to generate posterior distributions for the equation of state with filters accounting for both our present low-density nuclear physics knowledge and high-density neutron-star physics constraints.
The global structure of neutron stars thus predicted is discussed in connection with recent astrophysical observations.
\end{abstract}

\section{Introduction}

Born from core-collapse supernovae, neutron stars (NSs) are among the most compact objects in the Universe, with central densities that can reach several times nuclear saturation density, $n_{\rm sat} \approx 0.16$~fm$^{-3}$ 
\cite{Haensel2007}.
Below a thin atmosphere, the interior of a cold NS is expected to be composed of 
(i) an outer crust, 
made of fully ionised atoms arranged in a Coulomb lattice and neutralised by a degenerate electron gas,
(ii) an inner crust, 
where clusters and electrons coexist with unbound neutrons (so-called pasta phases may also be present at the bottom of the crust), 
(iii) a core, above $\sim 0.5 n_{\rm sat}$, usually divided into an outer core, made of a neutron-rich liquid with a small admixture of protons and leptons (electrons and eventually muons), and an inner core, formed of strongly interacting matter whose nature still remains largely unknown \cite{Haensel2007}.

From the observational point of view, enormous progress has been made by multi-messenger astronomy, providing different quantitative measurements of various NS properties, like masses (see \cite{Suleiman2021} for a compilation), radii \cite{Riley, Miller}, and tidal polarizability from NS-merger gravitational-wave data \cite{Abbott2018}. 
The connection between astrophysical observations and microphysical properties of NSs requires both microscopic and macroscopic (global simulation) modelling, where the specific microphysics inputs depend on the astrophysical scenario explored. 
If one restricts oneself to the study of cold (mature), non-rotating, unmagnetised, and isolated 
NSs, this connection mainly relies on the knowledge of the equation of state (EoS), i.e. the relation between the pressure $P$ and the mass-energy density of (baryonic) matter $\rho$.
Indeed, as far as static properties of these NSs are concerned, the hypothesis of validity of general relativity suffices to guarantee a one-to-one correspondence between such global properties (like the relation between the NS mass $M$ and radius $R$) and the corresponding EoS.
Thus, it would be in principle possible to reverse the problem and associate to simultaneous observations of mass and radius of several NSs the underlying $P$-$\rho$ relation, and possibly the corresponding composition.
However, the most precise mass measurements 
are available only for some NSs, mostly in binary systems, while measurements of radii, though becoming more precise, still suffer from model dependencies, making the constraints rather loose. 
On the other hand, additional complications arise from the EoS modelling, since the EoS itself is model-dependent; also, no ab-initio calculations exist in all regimes spanned by NSs, thus phenomenological models have to be employed instead \cite{Oertel2017, Burgio2018}.
Moreover, the relation between EoS and composition is not free from ambiguity, and very similar EoSs (thus almost indistinguishable $M$-$R$ relations) can be obtained under different hypotheses on the underlying microphysics, giving rise to the so-called ``masquerade'' effect \cite{Blaschke2018}.
At present, a number of NS EoSs are available (see e.g. the CompOSE database \cite{compose}), either with only nucleonic or with hyperonic and quark degrees of freedom. 
Most of them are non-unified, i.e. built piecewise, matching different models, each one applied to a specific NS region.
However, a thermodynamically consistent description of all regions of the NS is of utmost importance for (dynamical) simulations since the ad-hoc matching of different EoSs may trigger spurious instabilities and lead to uncertainties in the predictions \cite{Fortin2016, Suleiman2021}.

In this contribution, we briefly present the status on the NS EoS modelling in Sect.~\ref{sec:model} (here, we restrict ourselves to the discussion on purely nucleonic models), specifically in connection with constraints coming from nuclear physics. 
Using a meta-model approach, we also perform a Bayesian analysis and discuss the NS global properties thus obtained in Sect.~\ref{sec:results}, in comparison with both the outcome of some popular nuclear models and recent astrophysical observations. 

\section{Equation-of-state modelling}
\label{sec:model}

The theoretical description of matter in extreme conditions is a very challenging task. 
Current nuclear physics experiments cannot probe all the physical conditions encountered in NSs thus theoretical models are required to extrapolate to unknown regions. 
The undertaken theoretical approaches also depend on the relevant degrees of freedom of the problem and are usually divided into 
(i) ab-initio approaches, 
aiming at an exact solution of the nuclear many-body problem, particularly based on the use of chiral many-body perturbation theory with coupling constants fitted from few-nucleon ground-state data;
(ii) phenomenological approaches, relying on the density functional theory using effective interactions or functionals
which depend on a certain number of parameters fitted to reproduce properties of finite nuclei and nuclear matter. 
While the former models are usually restricted to homogeneous matter (i.e., the NS core), the latter ones are also applicable to the inhomogeneous crust \cite{Oertel2017, Burgio2018}. 
A different approach to construct the (phenomenological) EoS is to use purely agnostic (non-parametric) EoSs, that do not rely on any description of the nucleon-nucleon interaction but are subject to general physics constraints. 
These agnostic tools, widely used to extract the nuclear EoS from gravitational waveforms (see e.g. \cite{Landry2020} and references therein), are however not enough to pin down the NS properties; for example, 
they cannot predict the NS composition, a needed input, e.g., in NS cooling models.

A way to extract information on the structure and properties of dense matter is to use EoS parametrizations covering the parameter space of effective nuclear models, 
while allowing 
for the possibility of exploring different hypotheses on the effective degrees of freedom at high density.
This can be achieved by the so-called nucleonic meta-modelling \cite{Margueron2018}, i.e., by introducing a flexible energy functional able to reproduce existing effective nucleonic models, as well as interpolate between them. 
In particular, a Taylor expansion in $x = (n-n_{\rm sat})/(3 n_{\rm sat})$ up to order $N$ (here, $N=4$) around the saturation point ($n = n_{\rm sat}, \delta = (n_n-n_p)/n = 0$, with $n_{n(p)}$ the neutron (proton) density) allows the separation of the low-order derivatives, better determined by nuclear theory and experiments, from the high-order ones, having the highest uncertainties, 
\begin{equation}
e(n, \delta) \approx \sum_{k=0}^{N=4} \frac{1}{k!} \left( \left. \frac{d^k e_{\rm sat}}{dx^k} \right|_{x=0} + \left. \frac{d^k e_{\rm sym}}{dx^k} \right|_{x=0} \delta^2 \right) x^k \ .
\label{eq:metamodel}
\end{equation}
The coefficients of the expansion are directly related to the so-called empirical parameters, given by the successive derivatives of the energy functional at saturation.
In Eq.~(\ref{eq:metamodel}), $e_{\rm sat} = e(n,0)$ is the energy per baryon of symmetric $n_n=n_p$ matter and $e_{\rm sym} = e(n,1) - e(n,0)$ is the symmetry energy per baryon, defined here as the difference between the energy of pure neutron matter (NM) and that of symmetric matter. 
Using the common notation in the literature, we denote the symmetry energy coefficient $J = e_{\rm sym}(n=n_{\rm sat})$ and the slope of the symmetry energy $L = de_{\rm sym}/dx|_{x=0}$.
For realistic applications to NSs, the expansion describing the homogeneous nuclear matter in Eq.~(\ref{eq:metamodel}) was modified, introducing an extra density dependence induced by the effective mass and a low-density correction enforcing the zero-density limit \cite{Margueron2018}.
Moreover, to correctly model the NS, a specific treatment of the inhomogeneous crust is needed, such to ensure a unified EoS treatment. 
To this aim, a compressible liquid-drop (CLD) model approach for the ions was used, where the bulk energy was complemented with Coulomb, surface, and curvature contributions (see \cite{Carreau2019, Carreau2020a} for details).
The complete parameter set $\mathbf{X}$ of the model consists of 13 bulk parameters plus 5 surface parameters: for each set of the former ones, the latter parameters were consistently determined from a $\chi^2$-fit to the experimental masses from the Atomic Mass Evaluation \cite{AME2016}.
Given a nuclear model, i.e. a full set of $\mathbf{X}$ parameters, at each baryon density, the equilibrium composition and EoS were obtained variationally by minimising the energy density of the system under the condition of baryon number conservation, charge neutrality and beta equilibrium holding.
In the results presented in Sect.~\ref{sec:results}, only spherical clusters were considered in the crust; the properties of the pasta phase and the uncertainties in the pasta observables 
were discussed within the same framework in \cite{Dinh2021a, Dinh2021b}.
Within this approach, it is possible to incorporate our current nuclear-physics knowledge (see Sect.~\ref{sec:constraints}), and predict NS observables with controlled uncertainties, as presented in Sect.~\ref{sec:results}.

\subsection{Constraints on the equation of state from nuclear physics}
\label{sec:constraints}

\begin{figure}[ht]
\begin{minipage}[b]{18pc}
\centering
\vspace{0pt}
\includegraphics[width=\textwidth]{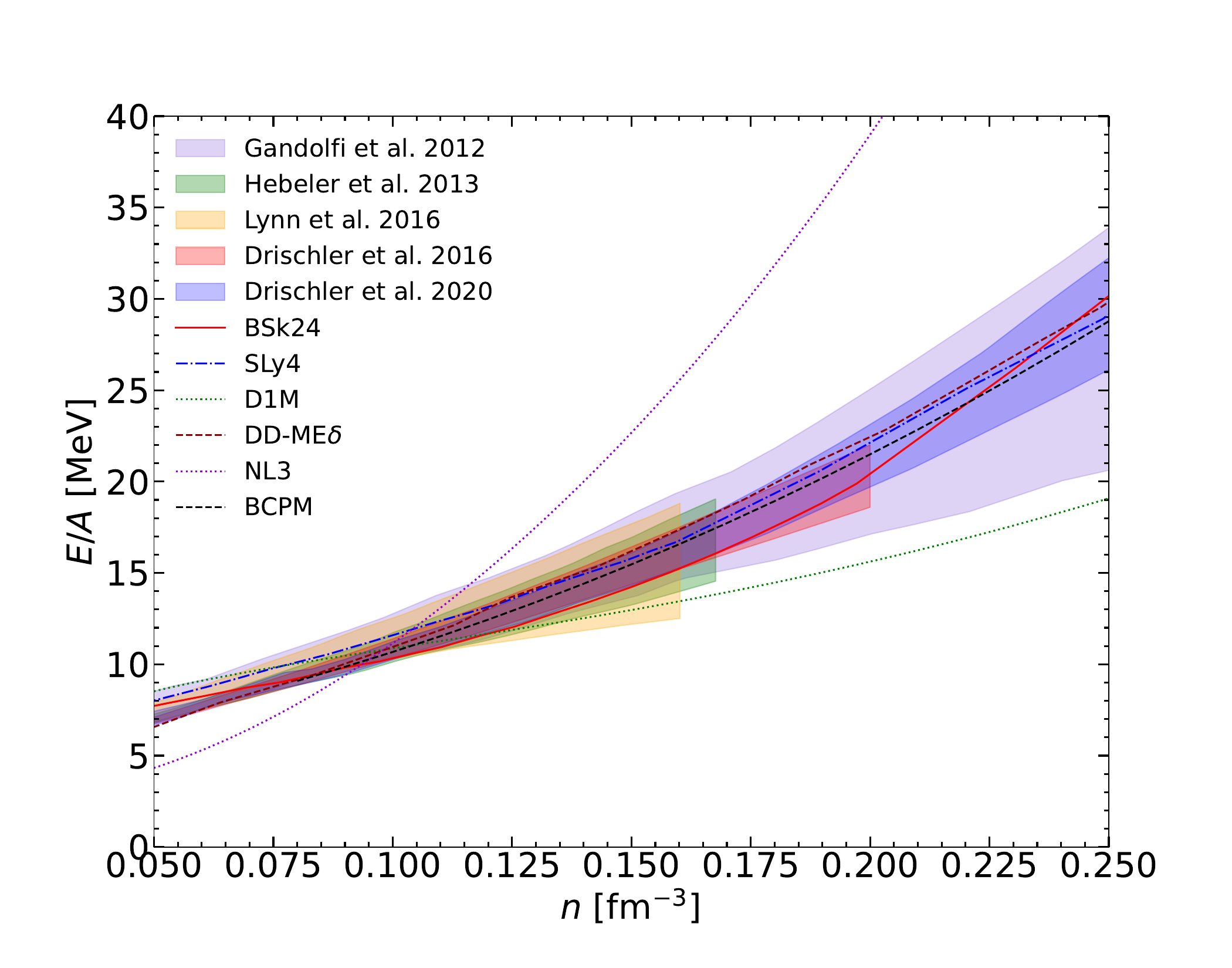}
\caption{Neutron matter energy per baryon versus baryon density for different calculations. 
See text for details.}
\label{fig:PNM}
\end{minipage}
\hspace{0pc}
\begin{minipage}[b]{18pc}
\centering
\vspace{0pt}
\includegraphics[width=0.82\textwidth]{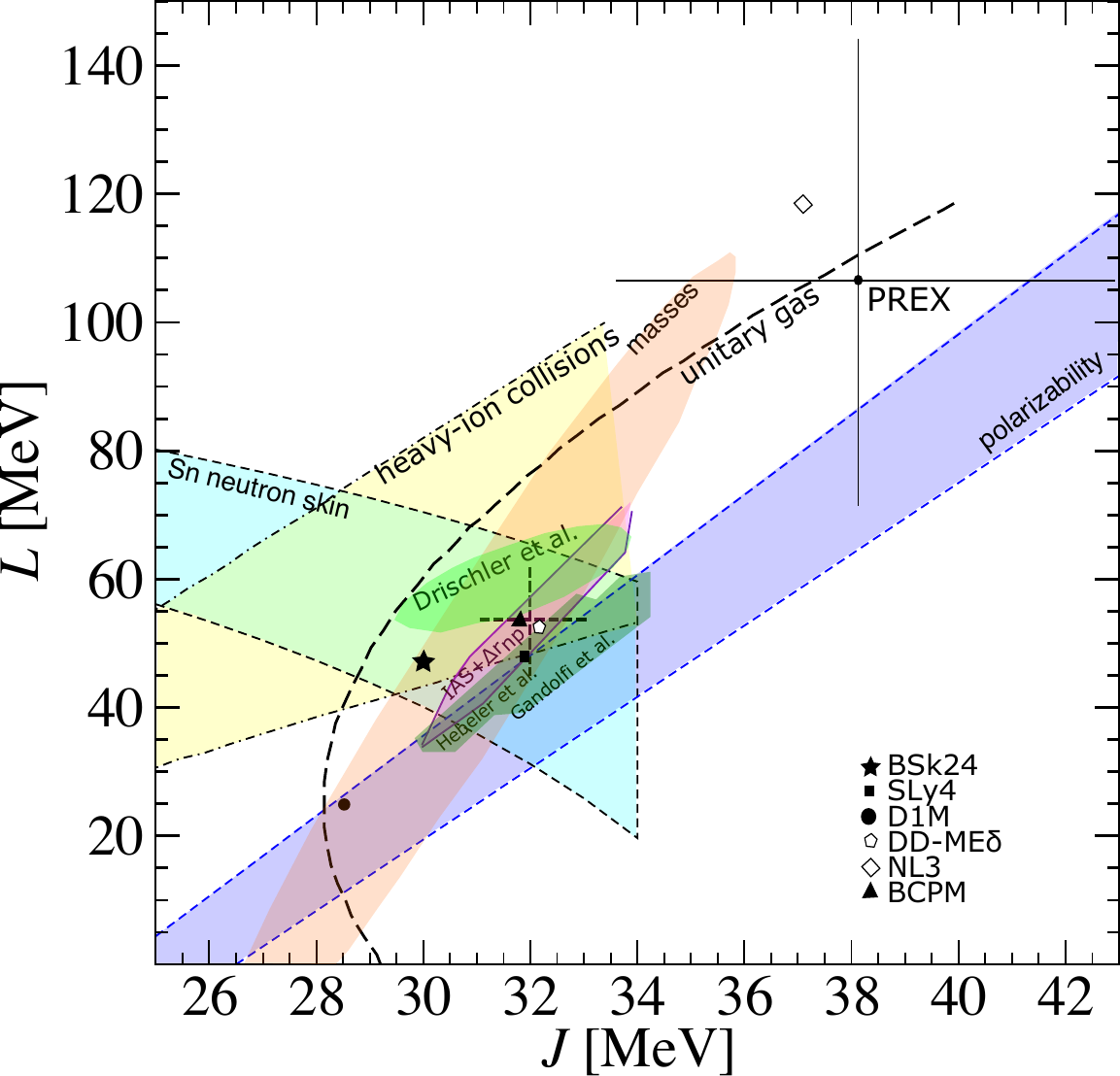}
\caption{Slope of the symmetry energy versus symmetry
energy coefficient extracted from different constraints. 
See text for details.}
\label{fig:J-L}
\end{minipage} 
\end{figure}

Theoretical models for the EoS can be constrained by both nuclear physics and astrophysics \cite{Oertel2017, Burgio2018}. 
The low-order empirical parameters 
are relatively well constrained by ab-initio nuclear theory and low-energy nuclear experiments. 
On the theory side, enormous progress has been made in ab-initio approaches, like the chiral effective field theory ($\chi$EFT), that is now able to quantify correlated uncertainties on the infinite-matter EoS. 
These models can provide powerful constraints on phenomenological models, particularly concerning predictions for pure NM.
This is illustrated in Fig.~\ref{fig:PNM}, where the NM energy per particle from different many-body calculations \cite{Gandolfi2012, Hebeler2013, Lynn2016, Drischler2016, Drischler2020} is plotted (shaded areas), showing a good agreement among them up to $\sim 1.5 n_{\rm sat}$.
For comparison, the predictions of some nuclear models used in astrophysical applications \cite{Goriely2013, SLy4, D1M, DDMEd, NL3, BCPM} are also displayed (data taken from \cite{Burgio2018, Goriely2013}).

On the experimental side, the low-order empirical parameters have been extracted by a systematic comparison of density functional approaches with a large set of nuclear observables, such as flows in heavy-ion collisions, nuclear masses, skins, isobaric analog states (IAS) systematics, electric dipole polarizability, and collective modes (see e.g. \cite{Burgio2018, Margueron2018, RocaMaza2018} for a review).
A (non-exhaustive) compilation of the constraints on the symmetry energy coefficient $J$ and slope $L$ from different experiments \cite{Tsang2009, Kortel2010, Chen2010, Daniel2014, RocaMaza2015} is shown in Fig.~\ref{fig:J-L}.
For comparison, constraints from ab-initio calculations are shown by green shaded areas \cite{Drischler2020}, while the dashed curve is the unitary gas limit on $(J,L)$ \cite{Tews2017}, meaning that values of $(J,L)$ to the right of the curve are permitted.
The predictions of $(J,L)$ from the nuclear models considered in Fig.~\ref{fig:PNM} are shown by symbols.
The tighter constraints are provided by nuclear theory, although there is no area of the parameter space where all the considered constraints are simultaneously fulfilled. 
This is likely due to the current uncertainties in the experimental measurements and to the model dependencies that plague the extraction of the constraints from the raw data; for the theory, the uncertainties may be due to the different many-body methods employed.
Also shown in Fig.~\ref{fig:J-L} is the recent PREX measurement \cite{Reed2021}, that points towards rather high values of $(J,L)$, overestimating other current limits. 
However, the estimate obtained in \cite{Reed2021} is not model independent, and systematic errors are so large that no clear conclusion can be drawn at this stage.
Incidentally, a recent analysis \cite{Reinhard2021}, performed using a different set of energy density functionals with respect to \cite{Reed2021}, yields much lower values for $(J,L)$ (dashed cross in Fig.~\ref{fig:J-L}), in agreement with other previous estimates.

From Figs.~\ref{fig:PNM} and \ref{fig:J-L}, we can deduce that ab-initio theory can give tighter constraints than experiments on the NM EoS.
This is because the latter ones mainly probe nuclei up to the drip-line, while extremely neutron-rich nuclei are not yet accessible.
On the other hand, the combined analysis of different nuclear experiments yields higher precision on the empirical parameters characterising symmetric matter (see e.g. the discussion in \cite{Oertel2017, Burgio2018} and references therein).

\section{Neutron-star equation of state and global properties}
\label{sec:results}

The different constraints imposed by theoretical calculations and nuclear-physics experiments can be incorporated in the meta-modelling approach described in Sect.~\ref{sec:model}, using Bayesian inference (see e.g. \cite{Carreau2019, Dinh2021a}).
To this aim, the model parameters were largely varied using flat non-informative priors, and posterior distributions were generated applying filters accounting for both our present low-density (LD) nuclear-physics knowledge and high-density (HD) constraints coming from general and NS physics (see \cite{Carreau2019,Dinh2021b}):
\begin{equation}
    p_{\rm post}(\mathbf{X}) = \mathcal{N} w_{\rm LD}(\mathbf{X}) w_{\rm HD}(\mathbf{X}) \exp\left(-\chi^2(\mathbf{X})/2\right) p_{\rm prior}(\mathbf{X}) \ ,
    \label{eq:bayes}
\end{equation}
with $\mathcal{N}$ the normalisation. 
The $w_{\rm LD}$ filter on the bulk parameters is given by the uncertainty band of the $\chi$EFT calculations of pure NM by \cite{Drischler2016} (see Fig.~\ref{fig:PNM}), which is interpreted as a 90\% confidence interval in $n \in [0.02-0.2]$~fm$^{-3}$. 
The $w_{\rm HD}$ filter is defined by imposing stability, 
causality, a positive symmetry energy at all densities, and the resulting EoS to support $M_{\rm max} > 1.97M_\odot$, where $M_{\rm max}$ is the maximum NS mass 
determined from the solution of the Tolmann-Oppenheimer-Volkoff equations \cite{Haensel2007} ($M_\odot$ being the solar mass).
The exponential term in Eq.~(\ref{eq:bayes}) quantifies the quality of experimental nuclear mass \cite{AME2016} reproduction of each set $\mathbf{X}$.

\begin{figure}[ht]
\begin{minipage}[t]{18pc}
\centering
\vspace{0pt}
\includegraphics[width=\textwidth]{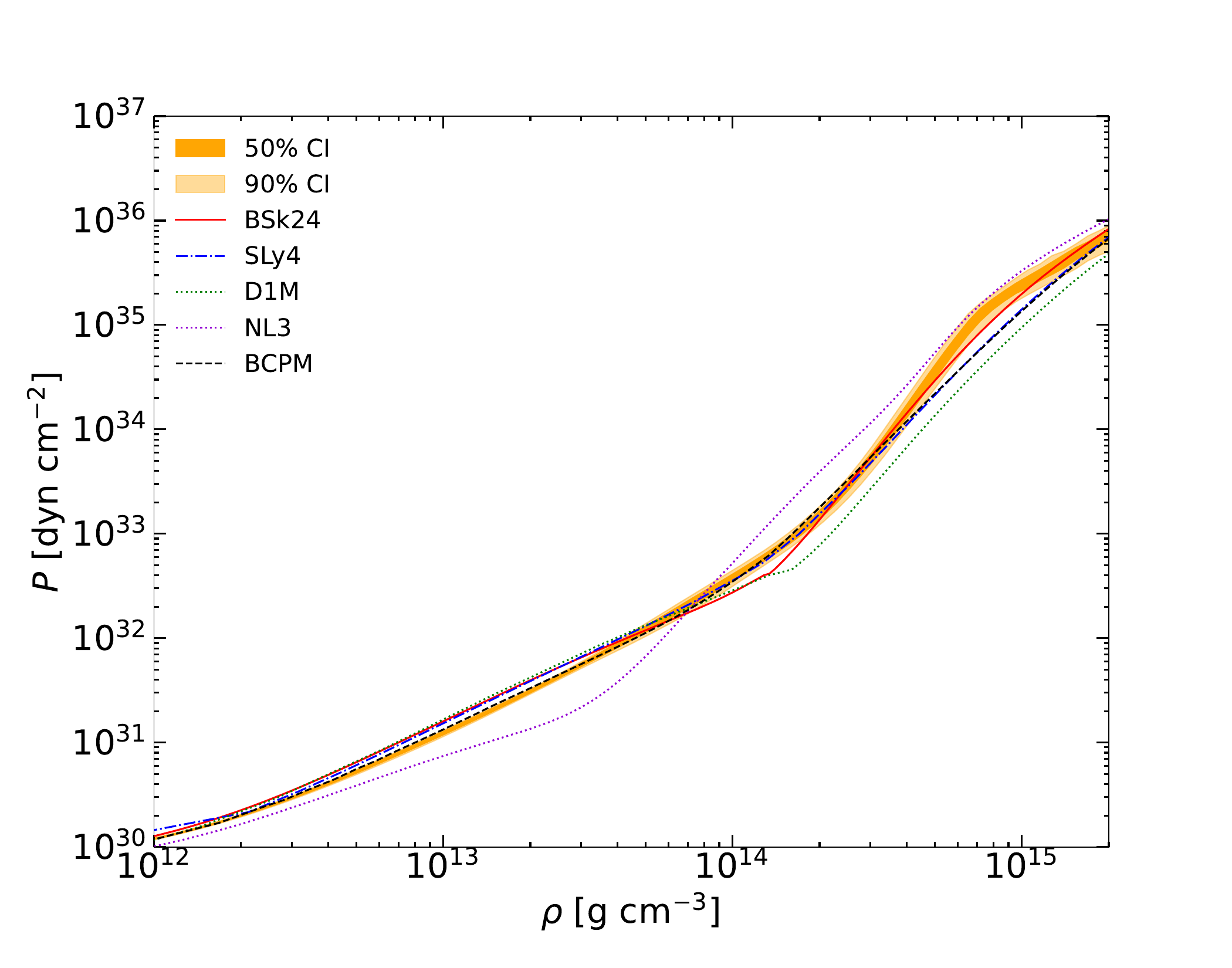}
\caption{Marginalised posterior for the NS EoS. 
See text for details.}
\label{fig:eos-NS}
\end{minipage}
\hspace{0.2pc}
\begin{minipage}[t]{18pc}
\centering
\vspace{0pt}
\includegraphics[width=\textwidth]{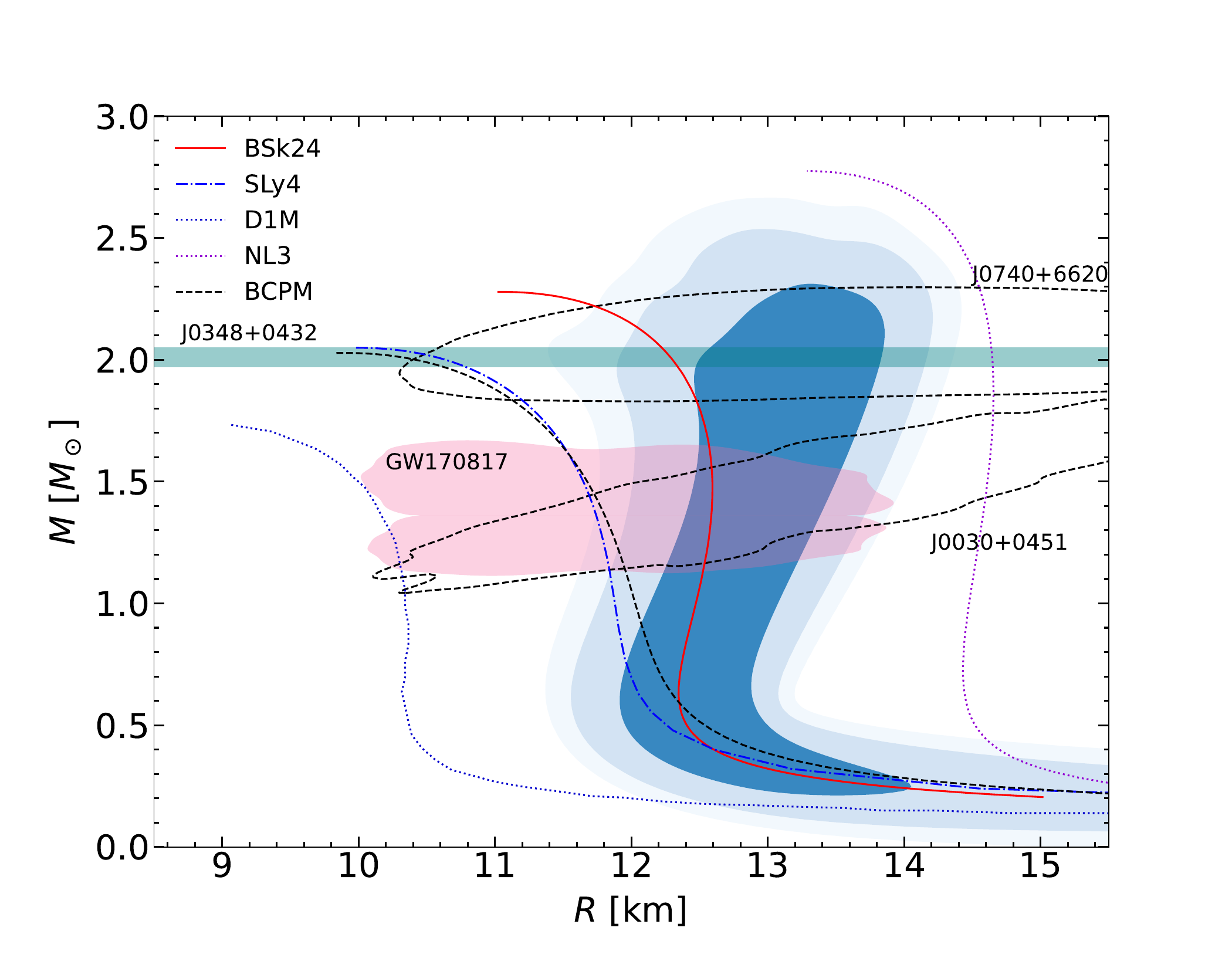}
\caption{Marginalised posterior for the NS mass-radius relation. 
See text for details.}
\label{fig:M-R}
\end{minipage} 
\end{figure}

The resulting posterior distribution for the NS EoS is shown in Fig.~\ref{fig:eos-NS}, the dark (light) yellow area corresponding to the $50\%$ ($90\%$) confidence interval.
For comparison, (unified) EoSs based on some models considered in Sect.~\ref{sec:constraints} are shown (data taken from \cite{compose, DH2001, BCPM, Pearson2018, Vinas2021, Xia2022}).
The EoS is quite tightly constrained at low density, because of the narrow LD filter, while a greater spread is noticed at high density, where the uncertainties on the nuclear functional are larger and the (less stringent) HD filters play a major role.
The corresponding NS mass-radius relation is shown in Fig.~\ref{fig:M-R}; the blue shaded areas (from dark to light) show the $1\sigma$, $2\sigma$, and $3\sigma$ confidence intervals.
For comparison, the $M-R$ relation obtained using the EoS models illustrated in Fig.~\ref{fig:eos-NS} are displayed (data taken from \cite{compose, DH2001, BCPM, Pearson2018, Vinas2021, Xia2022}). 
The black dashed contours (pink shaded area) represent the constraints inferred from NICER observations of the NSs J0740$+$6620 and J0030$+$0451 \cite{Miller} (from GW170817 \cite{Abbott2018}) at $2\sigma$;
the horizontal band corresponds to the precisely measured mass of the pulsar J0348$+$0432 \cite{Antoniadis2013}. 
As can be seen, our posterior, as well as the curves obtained using the EoSs based on the BSk24, SLy4, and BCPM functionals, are compatible with the shown astrophysical constraints, while some other popular models are not.
This is probably because most nuclear models have been optimised to reproduce specific nuclear data, where a limited range of densities and isospin asymmetries are explored. 
Thus, their extrapolation to beta-equilibrated neutron-rich matter in a large density domain as found in NSs might be incompatible with the combined set of constraints.
Incidentally, as noted in \cite{Dinh2021a}, the NL3 model, that predicts $(J,L)$ values close to those reported in \cite{Reed2021}, appears instead in disagreement with both ab-initio calculations and recent astrophysical observations (see Figs.~\ref{fig:PNM} and \ref{fig:M-R}).

Finally, we remind that the hypothesis underlying the employed meta-modelling is that only nucleons and leptons are present in the NS core, which might be restrictive at high densities in NSs.
However, this model can be also used as a null hypothesis to infer from the astrophysical data the presence of additional non-nucleonic degrees of freedom (see the discussion in \cite{Gulminelli2021}).

\section{Conclusions}

In this contribution, we have briefly presented the current status of the (NS) EoS modelling, in connection with present constraints from nuclear physics. 
Employing a nucleonic meta-modelling approach and performing a Bayesian analysis, we have discussed the resulting NS properties, namely the mass-radius relation.
Since we have assumed 
that only nucleons and leptons are present in the NS core, the question arises concerning the possible existence of additional degrees of freedom.
From the obtained posterior distribution, we can infer that, at present, observational constraints are not stringent enough to pin down the question, but future astronomical observations expected from current and future generation of (gravitational-wave) detectors could provide more compelling information on the structure and composition of NSs.

On the other hand, additional constraints on the EoS (particularly in what concerns the lower-order empirical parameters) could be provided by nuclear physics experiments.
These latter include more precise neutron-skin measurements and heavy-ion collision experiments, 
that can probe supra-saturation densities up to about $2 n_{\rm sat}$ (e.g. \cite{Huth2022} and references therein), thus allowing to reduce the uncertainties in the extrapolation of the nuclear EoS at high density.


\ack
This work has been partially supported by the
IN2P3 Master Project NewMAC and the CNRS International Research Project (IRP) ``Origine des \'el\'ements lourds dans l'univers: Astres Compacts et Nucl\'eosynth\`ese (ACNu)''. We thank H. Dinh Thi for useful suggestions.


\end{document}